\documentclass[fleqn,10pt]{wlscirep}
\usepackage[utf8]{inputenc}
\usepackage[T1]{fontenc}
\usepackage{epsfig}
\definecolor{mypurple}{rgb}{0.7,0.3,0.8}
\definecolor{midgreen}{rgb}{0.1,0.6,0.3}

\usepackage{lineno}

\title{Modeling the interplay between  epidemics and regional socio-economics}

\author[1,*]{Jan E. Snellman}
\author[2]{Rafael A. Barrio}
\author[1,3]{Kimmo K. Kaski}
\author[4,1,5]{Maarit J. Käpylä}
\affil[1]{Department of Computer Science, Aalto University School of Science, FI-00076 AALTO, Finland}
\affil[2]{Instituto de F{\'i}sica, Universidad Nacional Aut{\'o}noma de  M{\'e}xico, 01000 M{\'e}xico D.F., Mexico}
\affil[3]{The Alan Turing Institute, 96 Euston Rd, Kings Cross, London NW1 2DB, UK}
\affil[4]{Max-Planck-Institut f\"ur Sonnensystemforschung, Justus-von-Liebig-Weg 3, D-37077 G\"ottingen, Germany}
\affil[5]{Nordita, KTH Royal Institute of Technology \& Stockholm University, Hannes Alfv\'ens v\"ag 12, SE-11419, Sweden}

\affil[*]{jan.snellman@aalto.fi}

\begin{abstract}
In this study we present a dynamical agent-based model to investigate the interplay between the socio-economy of and SEIRS-type epidemic spreading over a geographical area, divided to smaller area districts and further to smallest area cells. The model treats the populations of cells and authorities of districts as agents, such that the former can reduce their economic activity and the latter can recommend economic activity reduction both with the overall goal to slow down the epidemic spreading. The agents make decisions with the aim of attaining as high socio-economic standings as possible relative to other agents of the same type by evaluating their standings based on the local and regional infection rates, compliance to the authorities' regulations, regional drops in economic activity, and  efforts to mitigate the spread of  epidemic. We find that the willingness of population  to comply with authorities' recommendations has the most drastic effect on the epidemic spreading: periodic waves spread almost unimpeded in non-compliant populations, while in compliant ones the spread is minimal with chaotic spreading pattern and significantly lower infection rates. Health and economic concerns of agents turn out to have lesser roles, the former increasing their efforts and the latter decreasing them.
\end{abstract}
\begin{document}

\flushbottom
\maketitle
%
%
\thispagestyle{empty}

\section*{Introduction}

In today's world variably populated urban and rural areas and people moving around and between them, are the main causes for infectious diseases spreading and wide-spread epidemic. Such wide-spread epidemics can cause health and economic concerns among general public, governments and local authorities as to what kind of restrictive measures should be taken to confine the epidemic spreading and mitigate its adverse socio-economic effects. How effective these non-pharmaceutical interventions (NPIs) turn out to be, depends to large extent, on the compliance of individuals. Thus the epidemic spread and the behaviour of people are intertwined, as the individuals' moves around and interactions with other individuals serve as the main engines of disease spreading. Apart from the NPIs by the authorities of a country or its districts, there is yet another issue for them to consider, namely what effects these measures have on the economy and welfare of the society. In order to understand the epidemic spreading and its socio-economic effects on a society of a country, we need to consider dynamical interplay of mutually intertwined disease spreading, disease carrying population acting on NPIs, and authorities implementing NPIs while taking care of social and economic welfare. 

To describe the dynamics of a complex system involving processes at different length and time scales, modelling is the main methodological approach. In a recent paper modelling the epidemic spreading itself, following four categories were identified \cite{PERRA2021}: 
(i) {\em{compartmental models}} with population divided to compartments reflecting its health status, (ii) {\em{metapopulation models}} based on networks of sub-populations connected by mobility, (iii) {\em{statistical models}} capturing the evolution of epidemic by inferring key parameters and behaviour from data, and (iv) {\em{agent-based models}} capturing the spreading patterns at the level of single individuals. Of these models the compartmental approach has become popular and quite successful in predicting the course of epidemics, because it can be extended to take into account different epidemiological properties of the infection spreading. However, in the present study we need to go further and describe the intertwined process of disease spreading, behaviour of population, authorities' NPIs, and society level welfare. For doing this we will construct a hybrid model of compartmental epidemics, linked with population agents and authority agents, the building blocks of which we discuss next.

For the basic building block to describe epidemic spreading we choose an extension of the compartmental SIR (Susceptible, Infected, Recovered) model \cite{KK1991a,KK1991b,KK1991c}, namely the SEIRS (Susceptible, Exposed, Infected, Recovered, Susceptible) model \cite{KK1991a,KK1991b,KK1991c}.  This model was further extended by Barrio et al. to describe epidemic spreading within a mobile population distributed over the geographical area of a country \cite{BVGM2013,BKHAG2020,BGBB2020}, such that the area was divided into a two dimensional grid of cells,  each containing a SEIRS model that evolves in time according to the population density of the cell. The probability of transmission of infection from one cell to another neighboring cell is assumed to be stochastic and dependent on a mobility parameter of the population of that cell. Apart from this local mobility there can be longer range mobility from one cell to another non-neighbor cell by various means of transportation. Furthermore, we point out that contagions, i.e. the flow from compartment E to  compartment I, are considered proportional to the probability for having at least one contact, which depends on the population density and includes time delays to define the disease progression.

As the other building blocks of our hybrid model we introduce two types of agents, namely the population agents in each of the SEIRS model grid cell of the  geographical area of a country and the authority agents governing the population agents in the collections of grid cells that are considered to constitute districts of that country. These two types of agents are introduced to describe the interactions between the popular and authorities' responses to epidemics and the epidemics itself, from the socio-economic or welfare perspective. The behaviour of both of these agents is based on a single assumption of status maximization as the source of human motivations, called Better Than Hypothesis (BTH), developed in previous studies \cite{SGGBK2017,SGKBK2018,SBK2021}. The BTH shares its basic assumption about the human motivation as being rooted in human psychology to the behavioural concepts of superiority and inferiority \cite{adler}, with social sciences to the desire for gaining status \cite{AHH2015}, and with neurological studies to how human brain processes status-relevant information \cite{ZTCB2008,ISS2008,KMD2012,UP2014}. Therefore, it makes sense to use an approach like the BTH in the context of comprehensive agent-based social simulations, given the ubiquity of social hierarchies in human societies.

In the context of epidemics, agent-based models have been employed to study the economic effects of societal closing or lockdown propagating through supply chains, \cite{IT2020,IMT2021}, and the impact 
of epidemic on economy \cite{CLV2020}. There are also a couple of agent-based studies to examine the connection between economy and public health \cite{BDHHK2020,SBLAGS2020,KSJJ2021}. In these studies the observed economic decline is caused by factors such as morbidity instead of the agents making conscious decisions in trying to protect themselves from the disease by, e.g., reducing their risk for getting infected in situations like shopping, public gatherings or other similar activities. Naturally, the questions of the psychological impacts of pandemic on both individual and societal levels are complex, making them challenging to model, but the BTH-approach allows systematic inclusion of different psychological effects to agent-based models. The public health vs. economy discourse concerns, in essence, societal values, thus raising the question which one to prioritize over the other? This question could be rephrased as "is it better to safeguard one at the expense of the other?", from which one could see how BTH relates to the topic of this study. However, there is a deeper connection between the BTH and societal values that stems from the simple fact that different people use different criteria to assess their overall social positions. In the context of a epidemic people may, for example, value their personal freedom over the authorities recommendations, and the authorities may value economic growth over mitigating the epidemic. Using the BTH one can model the effects of these attitudes on the behaviour of different human groups and, in the epidemic context, how this behaviour affects the spreading patterns of the disease.

This paper is organised such that in Section II we present our hybrid model, followed by the presentation of Results in Section III, and finally in Section IV Discussions the implications of our findings and applications to other contexts are discussed.

\section*{Hybrid socio-economy and epidemic spreading model}
\label{modsec}

\begin{figure}[ht]
\centering
\includegraphics[width=0.85\textwidth]{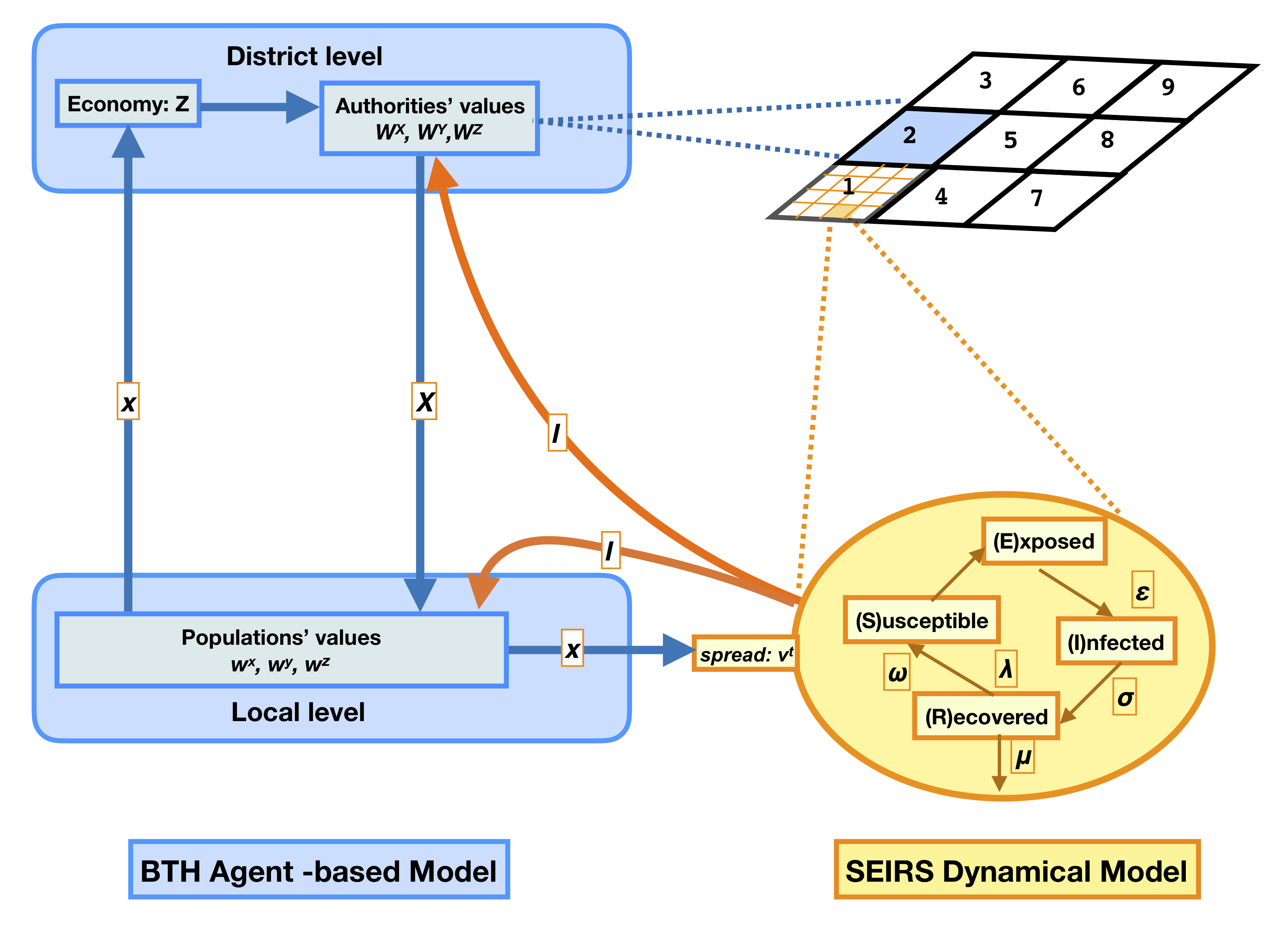}
\caption{
The flowchart of hybrid BTH agent-based socio-economy (blue squares) and SEIRS-based epidemic spreading (yellow ellipse) model. The BTH agent-based socio-economy consist of dynamics of local-level population agents (lower square) and district-level authority agents (upper square) embedded in the  rectangular geographical area of a "country" (upper right corner), divided to 9 districts (smaller squares, numbered) 
each with an authority agent and further to grid cells (smallest squares) each with a population agent. The thick blue and brown arrows illustrate the couplings between different parts of the model. The brown ones indicate information ($I$) of the state of epidemic spreading on which the population and districts agents react by reducing their economic activities ($x$) and ($X$) as indicated by the vertical blue arrows, respectively. The economic activity reduction by a population agent of a cell reduces the epidemic spreading ($v_i^t$) to neighbouring cells (the horizontal blue arrow). The parameters for the SEIRS and BTH-agents dynamics are described in the text.}
\label{scheme}
\end{figure}

In order to investigate the interactions between the responses of the population and authorities to the epidemics and their interplay with economics, we devise a geographical human society model of compartmental epidemics and socio-economies of local population and authority agents. As depicted in Fig.~\ref{scheme} the model consists of a rectangular geographical area of a "country" divided into several smaller districts (part of the grid coloured blue), with their own authorities and socio-economy. These districts are further subdivided into grids of geographical cells (part of the grid coloured yellow), each with its own population. The epidemic spreading takes place among the population of each cell as well as between the populations of different cells. In addition, the socio-economy of the geographical area is included in the model by introducing the behavioural dynamics of district authorities and geographical cell population agents. 

For describing the epidemic spreading we choose the SEIRS model introduced by Barrio et al. \cite{BVGM2013}. In this model, the disease can spread with probability $v^t$ cardinally from one cell to its nearest-neighbour ones in the grid of geographical cells or with some other probability to a more distant cell due to different means of transportation. However, for the sake of simplicity, such longer distance spreading is not considered in this study. 
As illustrated in Fig.~\ref{scheme} the SEIRS dynamics of each grid cell is governed by the following set of parameters: 
(i) $\epsilon$ is the period of latency before those who contract the disease become infectious, (ii) $\sigma$ the period of infectiousness,  (iii) $\omega$ the period of immunity, (iv) $\beta$ the transitivity of the disease, (v) $\mu$ the mortality rate, and (vi) $\lambda$ the portion of the population that is susceptible again after being recovered. The time-evolution of the epidemic, i.e. the numbers of susceptible (S), exposed (E), infected (I) and recovered (R), are characterized by the following set of difference equations:
\begin{linenomath*}
\begin{eqnarray}
S_{t+1} &=& q (S_t - G_t + \lambda q^b G_{t-1-b} ) + \mu N \nonumber \\
E_{t+1} &=& q (E_t + G_t - q^{\epsilon} G_{t-1-\epsilon} ) \nonumber \\
I_{t+1} &=& q (I_t + q^{\epsilon} G_{t-1-\epsilon}-q^a G_{t-1-a} ) \nonumber \\
R_{t+1} &=& q (R_t + q^a G_{t-1-a}-q^b G_{t-1-b} ),
\end{eqnarray}
\end{linenomath*}
where we denote $q = (1 - \mu$), $a = (\epsilon + \sigma$), $b = (\epsilon + \sigma + \omega$), and $G_{\alpha} = S_{\alpha} (1 - e^{-\beta I_{\alpha}})$. Whenever the epidemic spreads to a cell, the initial amount of infected is given by the parameter $\eta$, such that $S_{t_0} = 1 - \eta$ and $I_{t_0} = \eta$ at the time $t_0$ of spreading. In this study the time is measured in days and we fix the epidemic parameters to values used in the recent SEIRS studies of Covid-19 epidemics in a few countries  \cite{BKHAG2020,BGBB2020}: 
(i) the period of latency $\epsilon = 1$; (ii) the period of infectiousness $\sigma=14$; (iii) the period of immunity $\omega=140$; (iv) the transmission rate in cells $\beta=0.91$; (v) the mortality rate $\mu=3.69x10^{-5}$; (vi) the survival parameter $\lambda=0.9973$; (vii) the initial condition for the number of infected $\eta = 10^{-4}$. We note that, although these parameter choices seemed to give quite realistic results when applied to real countries, we will not attempt to adjust the values further, since the main purpose of the present study is to explore the behaviour of our hybrid model. 

The populations of the grid cells and authorities of districts of our socio-economic model are treated as agents, whose behavior is characterized by the choices they make to restrict their economic activity for lowering the chances of epidemic spreading. This requires a model of the dependence of the probability $v_i^t$ of epidemic spreading from one geographical cell to the next, on the economic activity. For the sake of simplicity, we assume
this relation to be linear:
\begin{linenomath*}
\begin{equation}
v_i^t = v_0 - x_i v_{max},
\label{vtdef}
\end{equation}
\end{linenomath*}
where $v_0$ is the maximum value of the mobility parameter, $0 \leq x_i \leq 1$ is the proportional reduction of the economic activity of the population agent $i$, and $v_{max} < v_0$ is the maximal reduction in the infectiousness of the disease that the economic measures can deliver. We assume that the population agents will not reduce their economic activity below the level necessary for their survival, meaning that $x_i$ refers specifically to the non-essential part of the economic activity of the agent. Here we choose the maximum epidemic spreading probability $v_0=0.5$, and the maximal reduction to this probability $v_{max}=0.49$, which means that the minimum possible spreading probability is $v_0=0.01$. 

On the other hand, as it comes to district authority agents, we assume that they set, based on the temporal state of epidemic, their recommendations for the minimum economic activity in the range, $0 \leq X_i \leq 1$, by restricting non-essential economic activities. The population agents react to these recommendations but also directly to the temporal state of epidemic. From now on, the quantities denoted by upper case letters refer to those belonging to the district authority agents, while the lower case letters refer to the quantities associated with the population agents. The regulations and restrictions, e.g. proximity rules and business closures, by the district authority agents affect the spending patterns of the population agents. For the sake of simplicity, we assume that the businesses of the geographical cells are thought to be incorporated to their corresponding population agents in the cells they operate. The model tracks the infection rate $y_i$, the reduction of non-essential economic activity $x_i$, and the degree of compliance $c_i$ of the population agent $i$, defined as the difference of the population agent's spending pattern and authority agent's
regulation $X^{r}_i$:
\begin{linenomath*}
\begin{equation}
c_i = x_i - X^{r}_i.
\label{compdef}
\end{equation}
\end{linenomath*}

The BTH is a generic method of constructing agent-based models based on the assumption of status maximization being the driving motivation of human beings. Agents are assumed to compare themselves to other associated agents according to appropriate measures or comparison scales, which naturally need to be chosen according to the situation BTH is being applied to. If an application of BTH includes more comparison scales than one, the different scales need to be weighted with agent-specific coefficients representing the values of the agents. Thus, the general form of the BTH utility function for a generic agent $i$ reads:
\begin{linenomath*}
\begin{equation}
\mathfrak{u}_i = \sum_{\mathfrak{a}} (\mathfrak{w}_i^{\mathfrak{a}} (\mathfrak{a}_i + \sum_{j \in \mathfrak{e}_i} (\mathfrak{a}_i - \mathfrak{a}_j))), 
\label{genueq}
\end{equation}
\end{linenomath*}
where $\mathfrak{a}_i$ is any measure by which agent $i$ can be compared to others, $\mathfrak{e}_i$ is the set of agents with which the agent $i$ compares itself, and $\mathfrak{w}_i^{\mathfrak{a}}$ are the weights that determine how important it is for the agent $i$ to rank high on the comparison scale $\mathfrak{a}$, i.e., how much value agent $i$ places on $\mathfrak{a}$. In this study we do not attempt to fit our model to any real situation, but consider the choices of the value parameters arbitrary at this point. 

In order to describe the behavior of population agents, we take into account the values that they associate with their economic contribution to fighting the virus ($x_i$), health (keeping local infection levels low), and compliance (with the authorities'  restrictions), the values of which are the weights $w^x_i$, $w^y_i$ and $w^c_i$, respectively. Then, the utility function of the population agents takes the following form

\begin{linenomath*}
\begin{equation}
u_i = w^x_i (x_i + \sum_{j \in e_i} (x_i - x_j)) 
 + w^y_i (y_i + \sum_{j \in e_i} (y_i - y_j)) 
 + w^c_i (c_i + \sum_{j \in e_i} (c_i - c_j)),
\label{Ueqpcomp}
\end{equation}
\end{linenomath*}
where $y_i$ and $y_j$ are the infection rates suffered by population agents $i$ and $j$. It is assumed for simplicity that the populations put minimum effort to fighting the epidemic, i.e. choosing $x_i$ in such a way that $u_i = 0$. It should be noted that this choice implies that $w^x_i < 0$ for all the population agents. This assumption allows us to solve $x_i$ easily using Eq.~(\ref{Ueqpcomp})

\begin{linenomath*}
\begin{equation}
x_i = \frac{1}{|e_i| + 1} ( \frac{w^x_i}{w^x_i + w^c_i} \sum_{j \in e_i} x_j  
 - \frac{w^y_i}{w^x_i + w^c_i} (y_i + \sum_{j \in e_i} (y_i - y_j))) 
 + \frac{w^c_i}{w^x_i + w^c_i} (|e_i| X^r_i - \sum_{j \in e_i} c_j), 
\label{compUeq}
\end{equation}
\end{linenomath*}
where $|\cdot|$-notation refers to the number of members in a group. Another noteworthy point is that, to avoid division by zero, the values of $w^x_i$ and $w^c_i$ can not cancel each other out, i.e. $w^x_i + w^c_i \neq 0$. The meaning of this feature becomes evident from Eq.~\ref{Ueqpcomp}: if one inserts the definition \ref{compdef} to Eq.~\ref{Ueqpcomp}, one finds that all the $x_i$:s cancel each other out if $w^x_i + w^c_i = 0$, with the result that the actions of the population agents have no effect on their utility. This could be characterised as complete indifference by the population agents towards their contributions to fighting the epidemic.  

Analogously to the population agents, the district authority agents value making efforts to keep the epidemic at bay, which in their case amounts to make recommendations or restrictions for the population agents to follow, while aiming at low infection rates. These values are defined by the weights $W^X_i$ and $W^Y_i$, respectively, for the district authority agent $i$. As the setters of restrictions or rules, they do not have the compliance characteristic of the population agents. Instead, we assume the authority agent of district $i$ being concerned of the overall economic activity (or gross domestic product (gdp)) $Z_i$ and its relative reduction due to restrictions, $Z'_i$, defined here as follows
\begin{linenomath*}
\begin{equation}
Z_i = 1 - Z'_i, \text{ with,  }
Z'_i = \frac{1}{|P_i|} \sum_{k \in P_i} x_k,
\label{gdpeq}
\end{equation}
\end{linenomath*}
where $P_i$ is the set of all population agents residing in the geographical area the district authority 
agent $i$ governs. 
The weight that describes the value the district authority agent attaches to $Z'_i$ is denoted with $W^Z_i$. With these choices the district authority agent's utility function takes the form,
\begin{linenomath*}
\begin{equation}
U_i = W^X_i (X_i + \sum_{j \in E_i} (X_i - X_j)) 
 + W^Y_i (Y_i + \sum_{j \in E_i} (Y_i - Y_j)) 
 + W^Z_i (Z'_i + \sum_{j \in E_i} (Z'_i - Z'_j)),
\label{govUeq}
\end{equation}
\end{linenomath*}
where $Y_i$ is the average infection rate of district $i$. Just like in the case of the population agents, it is assumed that the district authority agents want to keep the restrictions to the minimum, meaning that they choose their recommendations or restrictions $X_i$ so that $U_i = 0$. Again, this implies $W^X_i < 0$ for all the authority agents. Hence from Eq.~(\ref{govUeq}) we obtain: 

\begin{linenomath*}
\begin{equation}
X_i = \frac{1}{|E_i| + 1} (\sum_{j \in E_i} X_j - \frac{W^Y_i}{W^X_i} (Y_i + \sum_{j \in E_i} (Y_i - Y_j)) 
 - \frac{W^Z_i}{W^X_i} (Z_i + \sum_{j \in E_i} (Z_i - Z_j))).
\label{dmeqgovmkII}
\end{equation}
\end{linenomath*}

In Eqs.~\ref{compUeq} and \ref{dmeqgovmkII} we assume that if $x_i$ and $X_i$ get outside their range, they are set to  $0$ or $1$, 
depending on whether the result is below the former, or larger than the latter. One of the basic properties of these value variables is that their sign determines, whether the agent seek to minimize or maximize the quantity that the parameter refers to, while the absolute value of the parameter determines the strength of this tendency. Thus, for example, a district authority agent with negative $W^Y_i$ would seek to minimize the infection rate in its own district, whereas an authority agent with positive $W^Y_i$ would take the herd immunity approach and seek to encourage the spread of the epidemic. However, it should be noted that in deriving Eqs.~\ref{compUeq} and \ref{dmeqgovmkII} we have assumed that the agents will do the minimum allowed by their utilities, so studying the effects of a true herd immunity approach is beyond the scope of this study. Hence, we assume that $W^Y_i < 0$ for all the district authority agents, and likewise $w^y_i < 0$ for all the population agents.

Further, we assume that the district authority agents compare themselves to all other district authority agents in the model, while the population agents compare themselves with the nearest and next nearest neighbour agents. As we use non-reflecting wall boundary conditions, the number of neighbours for an agent to compare with, counts eight, five, and three in the middle, at the boundary and in the corner of the whole model area, respectively. In other words, $E_i$ contains all district  authority agents except the agent $i$, and $e_i$ contains the population agents that share a side or a corner with the population agent $i$. This means that $|E_i| = N_g - 1$, when the number of all the district authority agents is $N_g$, and $|e_i| = 8, 5$ or $3$ depending on the geographical location of the population agent $i$.  

Our hybrid model with SEIRS-epidemics and BTH agents of grid cell populations and district authorities are set on a large square-shaped geographical area or "country". As depicted in Fig.~\ref{scheme} this area is divided to a grid of $N$=51x51 cells, each of which has a BTH population agent and in each of which the epidemic follows the SEIRS-dynamics. This grid is grouped to larger 3x3 grid or $N_g=9$ jurisdiction or governance districts, numbered in Fig.~\ref{scheme}, each with 17x17 population cells for BTH-authority agents. The numbers of cells and districts were chosen odd to set the epidemic starting from the center of the geographical area and its district in the middle.

\section*{Results}
\label{resec}

\begin{figure}[ht!]
\centering
\includegraphics[width=0.40\columnwidth]{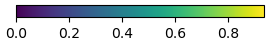}\\
\hspace*{-0.53cm}\epsfxsize = 0.435\columnwidth \epsffile{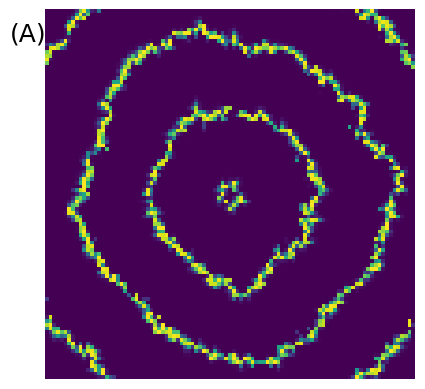}
\epsfxsize = 0.45\columnwidth \epsfysize = 0.3\textheight \epsffile{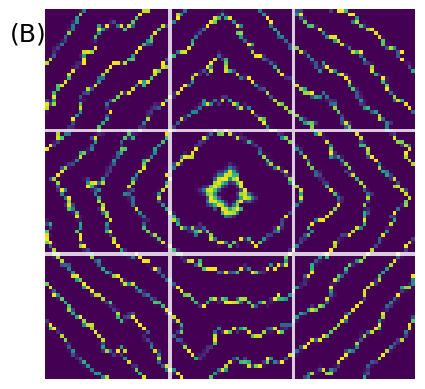}
\caption{(A) Snapshot of the epidemic spreading over a single region without agent activity.
The spreading probability, $v^t$, is $0.1$ everywhere. (B) Same as (A) but with agents' activities enabled, in which case $v^t$ changes according to Eq.~(3). We vary the value parameters 
for population and authority agents in
nine different districts.}
\label{lcontrol}
\end{figure}

In this section we present the results for our hybrid model based on SEIRS epidemic spreading and BTH agents of district authorities and  grid cell populations, which  we call the BTH-SEIRS model or just the hybrid model. We start by conducting control runs of the pure SEIRS model with the fixed parameter values set above. These results are then compared with those of the BTH-SEIRS model with 
variable BTH value parameters for all the agents (next section). Then we proceed by running the BTH-SEIRS model with different and fixed BTH value parameters to illustrate their effects and isolate the most important of them for the system dynamics. For the sake of simplicity, we set the parameter values the same for all the  authority agents and the population agents of the model. 

%

\subsection*{Comparison between pure SEIRS and BTH-SEIRS models}

In this subsection we investigate what the agent-based features add to the descriptive power of the the original SEIRS-model. Obviously some effect could be expected, as the addition of the BTH-agents introduces a connection between the epidemic and its socio-economic effects. The fact that the probability of the epidemic spreading to new areas is linked to the responses of the local population and district authority agents, brings in new geographical variability to the time evolution of the epidemic. In this section we demonstrate the basic properties of the epidemic spreading patterns in both the original SEIRS model and the BTH-SEIRS model with inhomogenous value parameters.

\subsubsection*{SEIRS model}

Since in the case of the pure SEIRS model the epidemic spreading probability $\nu^t$ is kept constant everywhere, it is straightforward to demonstrate different cases of spreading dynamics. It is easily seen that very high vales for $\nu^t$ causes the epidemic to propagate as wave fronts, while the low values of $\nu^t$ causes the epidemic to spread in a chaotic manner. As our simulation domain is a nearest-neighbour grid of square cells (see Fig.~\ref{scheme}) the epidemic spreads only in cardinal directions, and the wave fronts for $\nu^t = 1$ take the form of regular diagonal squares. When the spreading probability ($\nu^t$) is decreased from $1$, the wave fronts gradually lose their regularity, until all cohesion is lost and the spreading pattern appears random. This transition from the regular state to a chaotic one occurs when the spreading probability is set to a value below approximately $0.02$. In Fig. \ref{lcontrol}(A) we show the epidemic spreading when the spreading probability $v^t$ is set to a constant $0.1$, everywhere. In this case the epidemic spreads as wave fronts that resemble highly distorted square patterns. 

\subsubsection*{BTH-SEIRS model}
\label{BTHsec1}

For the BTH-SEIRS model, $\nu^{t}$ is no longer a constant, but varies spatially and over time. The parameter values for the district authority and population agents we choose at random from a uniform distribution of width one and centered around representative mean values as follows $-1.5 \leq W^x_i \leq -0.5$, $-10.5 \leq W^y_i \leq -9.5$, $-0.5 \leq W^z_i \leq 0.5$, $-1.5 \leq w^x_i \leq -0.5$, $-100.5 \leq w^y_i \leq -99.5$ and  $-0.5 \leq w^c_i \leq 0.5$.  The simulated spreading pattern for these parameter value choices is depicted in Fig. \ref{lcontrol}(B), where we find that in the vicinity of the center of the epidemic outbreak, the epidemic spreads in regular waves with the characteristic rectangular shape. This regular spreading pattern suggests that the agents did not take significant action to curtail the epidemic right at the start. Only slightly away from the centre, the spreading pattern becomes more irregular. This is because the actions of the local population agents, varying from time to time and place to place, become more pronounced, affecting strongly the spreading speed. There are also significant local variations in the spread of the epidemic both between and within the districts that are not present in the pure SEIRS model. 

\subsection*{BTH-SEIRS model: some specific cases}

In this subsection we have run numerous BTH-SEIRS model simulations 
to isolated the most important parameter values and discuss their effects. It turns out that the easiest way to make the simulated epidemic to exhibit either the random or wave-like spreading behaviour is to change the sign of the compliance $w^c_i$ of the population agents. We observe the epidemic spreading taking place as a wave front for non-compliant populations ($w^c_i < 0$), while simulation for the compliant populations ($w^c_i > 0$) the pattern is random. On the other hand increasing the concern for health of the population agents (i.e. $w^y_i$) causes the spreading patterns to fall in between these two extremes. For the sake of simplicity, all the simulations discussed in this section had constant value parameters for all the authority and population agents, unlike in the example simulation above. It should be noted that the value parameters are chosen entirely on the basis of how well they bring out the main types of behaviour or how well they demonstrate the effects of certain parameters.  

\subsubsection*{Effect of population agents' compliance}

As stated above, the extreme cases of spreading patterns of our model come to fore in situations where there is no significant variance in the parameter values of the population agents and, crucially, the signs of all the parameters are the same for all the agents. We found this out when we were investigating the effect of the compliance parameter $w^c_i$ of the population agents by setting the values of the district authority agents as follows: $W^X_i = W^Z_i = -1$, $W^Y_i = -10$. Here the infection parameter value $W^Y_i$ was set to be much lower than the other two parameters so that the authority agents would insist on relatively tough measures against the epidemic, which in turn will make the compliance more relevant. We have made two sets of runs, one with $w^x_i = w^y_i = -1$ and $w^c_i = -2$, and the other with $w^x_i = w^y_i = -1$ and $w^c_i = 2$, and name them the cases of Non-compliance and Compliance, respectively. Note that the absolute value of $2$ for $w^c_i$ was chosen to avoid division by zero in Eq. \ref{compUeq}. 

\begin{figure}[ht!]
\centering
\hspace*{-0.53cm}\epsfxsize = 0.45\columnwidth \epsffile{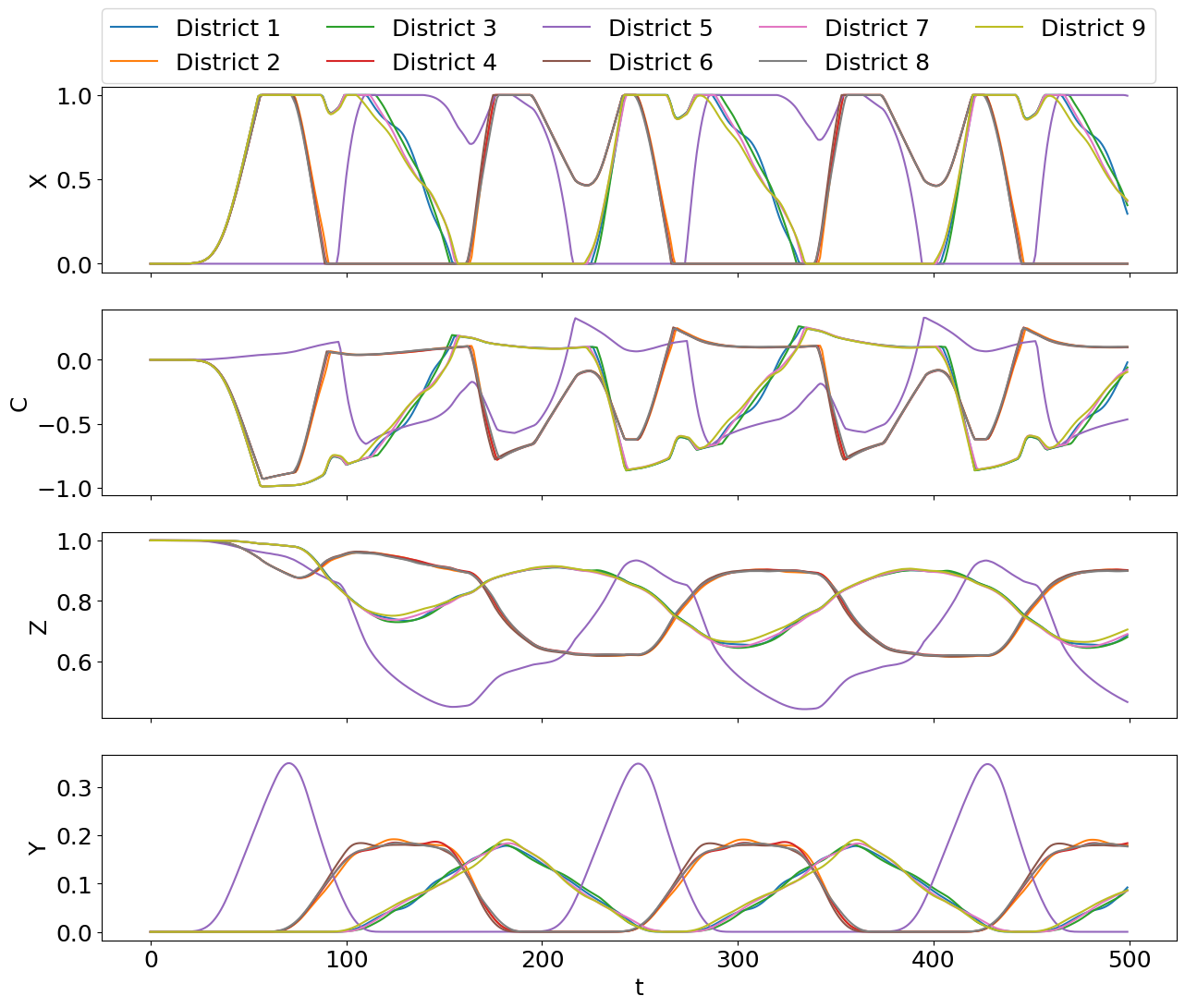}
\epsfxsize = 0.4\columnwidth \epsfysize = 0.3\textheight \epsffile{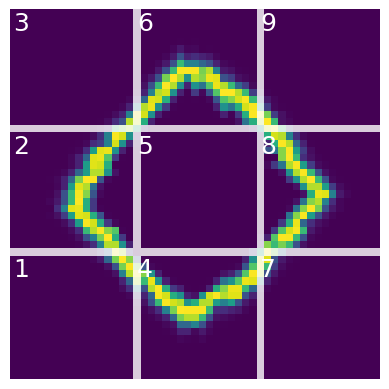}
\caption{The results for the Non-compliance case, i.e., simulating non-compliant populations. The four panels on the left show, from bottom to top, the average infection rate $Y_i$, economic activity or gdp $Z_i$, average degree of compliance $C_i$ of the population agents, and district authority agents' regulations $X_i$, while the panel on the right shows the spreading pattern of the epidemic over the geographical area of nine districts ($i=1, 2, 3, ...,9)$, numbered as in Fig. \ref{scheme}.}
\label{Cminus}
\end{figure}

The results for the Non-compliance case, in which the population agents' values are strongly against the district authorities' regulations, are shown in Fig. \ref{Cminus}. In the left panels (from bottom to top) we show district-wise the average infection rate $Y_i$, the economic activity or gdp ($Z_i$, see Eq. \ref{gdpeq}), the average degree of compliance of  population agents $C_i$, and the regulations of district authorities agents $X_i$ as functions of time. The right panel shows the characteristic spreading pattern when the first wave of the epidemic has reached all the districts. Here the spreading pattern appears to be of squarish-shaped with somewhat ragged edges, which is consistent with fast epidemic propagation speed, discussed above. Due to the rapid propagation and the relatively small size of the simulation domain, the later waves of the epidemic have not yet materialized. However, as the simulation progresses, new waves emerge periodically from the origin following the very same pattern as the first wave. The time period between the epidemic wave fronts is determined mostly by the length of immunity period $\omega$. This seasonal nature of the epidemic is clearly visible in the periodicity of the infection rates. Quite understandably, the behaviour of the different districts is strongly dependent on their distance from the origin of the epidemic, but the fact that equidistant areas show almost identical trends in every tracked quantity is remarkable. Especially the infection rates are almost the same for the districts in the cardinal directions ($2$, $4$, $6$ and $8$, hereafter called edge sharing districts) from the origin district ($5$), and the diagonally neighbouring districts ($1$, $3$, $7$ and $9$, hereafter corner sharing districts) likewise show  similar infection rates all the times. 
\begin{figure}[ht]
\centering
\hspace*{-0.53cm}\epsfxsize = 0.45\columnwidth \epsffile{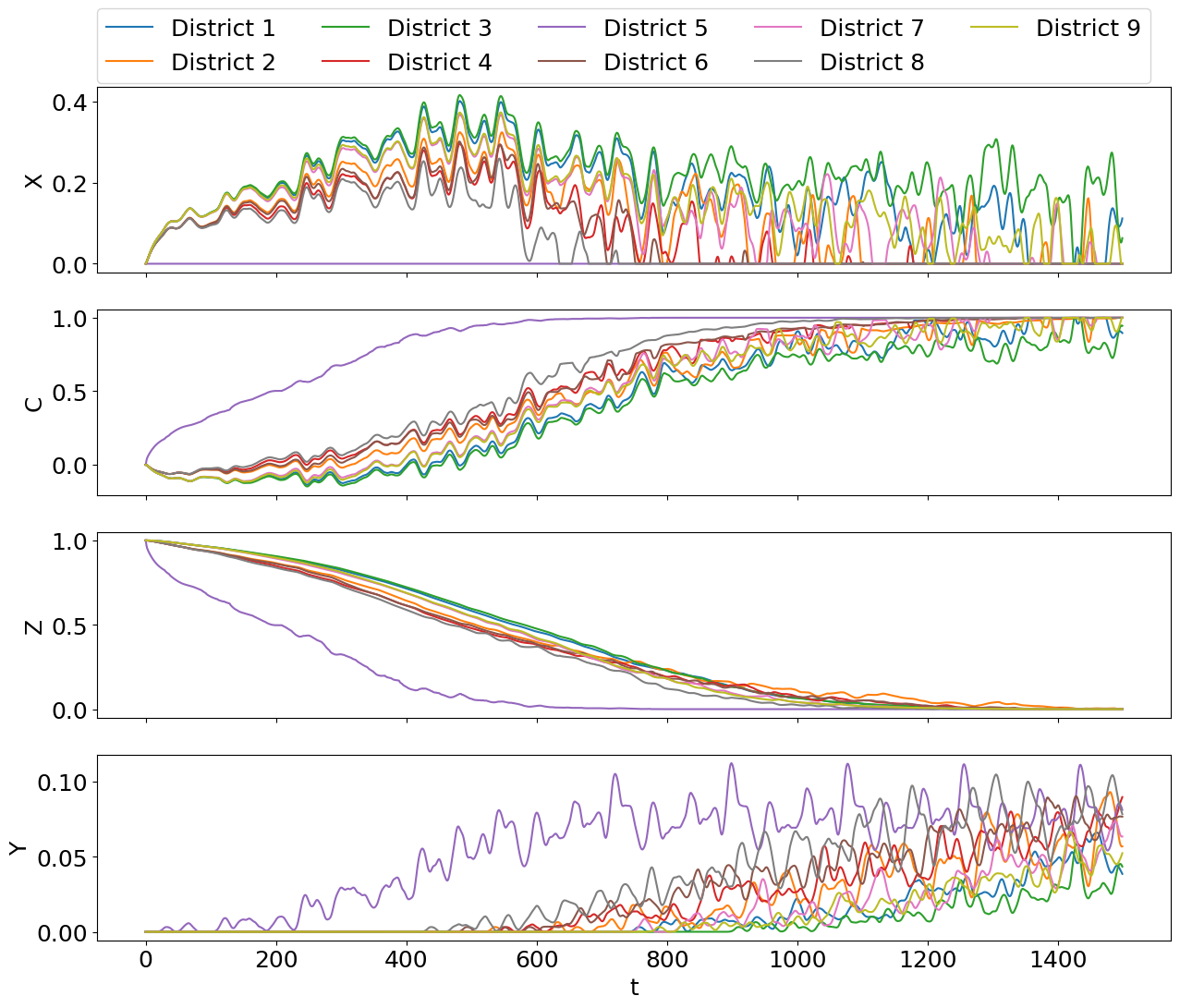}
\epsfxsize = 0.4\columnwidth \epsfysize = 0.3\textheight \epsffile{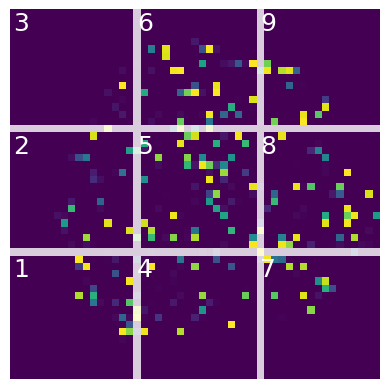}
\caption{The results for the Compliance case, i.e., simulating compliant populations. The notations are the same as in Fig. \ref{Cminus}.
}
\label{Cplus}
\end{figure}

An interesting detail that can be extracted from the infection rate curves is that their overall shapes can be deduced from the underlying epidemic spreading pattern. The central district (number 5) experiences a very steep rise and fall in its infection rate, because the epidemic spreads there in the full wave pattern, while other districts only get to experience the traveling wave fronts of the epidemic. Thus the peak infection rate of approximately $0.35$ in the central district is roughly twice that of the other districts showing  similar peak rates.

We note that the direction from which the wave front arrives plays a crucial role in determining its shape outside the central district. If the epidemic enters the edge sharing districts from the side, its evolution in these districts resembles a wedge travelling through them along the cardinal directions.  The result for the infection rate is that it reaches a peak value quite fast, but then linger there until the wedge of the epidemic front exits the district. This is seen in Fig. \ref{Cminus} as a flat top of the infection curves ($Y_i$). In contrast, in the corner sharing districts the epidemic enters from their corners and spreads as a front perpendicular to the diagonal direction. This causes the infection rates ($Y_i$) to climb and decline slower than in the other districts, since the peak occurs when the epidemic front runs diagonally from one corner of the district to the other.  

The general behaviour of district authority agents ($X_i$), as depicted in Fig. \ref{Cminus}, is characterised by periods of high economic restrictions reaching even the maximum level of $1$ and intervening periods with all the restrictions lifted up. Like for the infection rates, the patterns repeat themselves cyclically, although they do not fully settle until the second wave. There is also a similar correspondence between the actions of district authority agents controlling geographically similarly positioned districts, i.e.  authority agents of edge sharing districts follow the same trend as other edge sharing districts. The same holds for the corner sharing districts. However, while the policies of the authority agents of edge sharing districts are almost identical at all times, the authority agents of corner sharing districts have more variation among their policies when lifting restrictions. 

An interesting observation can be made by comparing the district authority agents' restrictions and infection dynamics. The district authority agents tend to impose restrictions before the epidemic arrives, but drop them as it takes hold, as if they were giving up after failing to contain the outbreak. Even more interestingly, the district $5$, where the epidemic starts, does not put up any restrictions initially, and is introducing them for the first time only after the adjacent areas have dropped theirs. The reason for this behaviour can be found in Eq. \ref{dmeqgovmkII}. If there are significantly more infections in the district governed by its authority agent $i$ than there are in the other districts the second term of this equation has an overall negative sign for the district authority agent $i$. Since $W^Y_i$ is much larger than $W^X_i$ and $W^Z_i$, the other terms cannot compensate for it, and so the regulations are set to remain at the minimum possible value of $0$ until the epidemic intensifies in the other districts. At this point the second term acquires positive sign and causes restrictions to be imposed. Moreover, the actions of the district authority agents influence each other, as dictated by the first term of Eq. \ref{dmeqgovmkII}. The introduction of restrictions by the central district coincides with the tightening restrictions in the corner sharing districts 
and the reintroduction of restrictions in the edge sharing districts similarly seems to encourage tighter restrictions in the central district.

The actions of the population agents, as seen in the economic activity (or gdp) ($Z_i$) and compliance degree ($C_i$) curves, are also very heavily determined by the district they belong to.  Again, the population agents of the edge sharing and corner sharing districts behave very similarly between themselves, while the population in the central district ($i = 5$) follows its own behavioural pattern. The degree of compliance ($C_i$) turns out to be low, about $0.25$ at best. As one would expect from the choice of the compliance value parameter, it is often negative, attaining values as low as $-1$. There is a clear correspondence between the tightening of the restrictions by the district authority agents ($X_i$) and the lowering of the degree of compliance ($C_i$). This is apparently driven primarily by extremely fast impositions of restrictions by the district authority agents, rather than by active resistance to these restrictions by the population agents. This is because we observe the economic activity ($Z_i$) curves also falling at the same time as the degree of compliance curve ($C_i$). However, the actions of population agents rarely meet the regulatory demands, and as a result the economic activity in any of the districts never falls to less than $40\%$ of the pre-epidemic levels. From the economic activity curve one can also see that the population agents tend to act to prevent the epidemic, and give up when they fail. This is evident from the fact that the deepest fall in the economic activity occurs immediately before the epidemic again picks up in a district. As the epidemic worsens, the population agents increase their spending, in accordance with how the district authority agents  dismantle the restrictions during the worsening phase of the epidemic. 

The results for the Compliance case are shown in Fig. \ref{Cplus}. In this simulation, the populations act decisively to halt the spread of the epidemic, which is clearly seen in the economic activity ($Z_i$) and degree of compliance ($C_i$) curves, as they tend to their minimum and and maximum values, respectively. This results in the epidemic spreading pattern being random and taking a long time to reach other areas outside the starting central district $i=5$. The authority agent of the central district ($i=5$) are not found to issue any restrictions at all, so the degree of compliance curve ($C_i$) for that district is a mirror image of the epidemic activity curve ($Z_i$). While all the district authority agents other than that of the central district ($i=5$) put in place some restrictions, they again tend to drop them some time before the epidemic enters their areas. As one would expect, the rule of thumb is that the edge sharing districts drop theirs first, and the corner sharing districts last, but due to the chaotic spreading of the epidemic this pattern is somewhat blurred. 

It should be noted that both the infection rates and the restrictions by the district authorities are much lower in this case than in the case of Non-compliance. Furthermore, it could be said that in this case the district authorities to some extent cede their responsibility for mitigating the spread of epidemic within the population of the district. 

\begin{figure}[ht]
\centering
\hspace*{-0.53cm}\epsfxsize = 0.45\columnwidth \epsffile{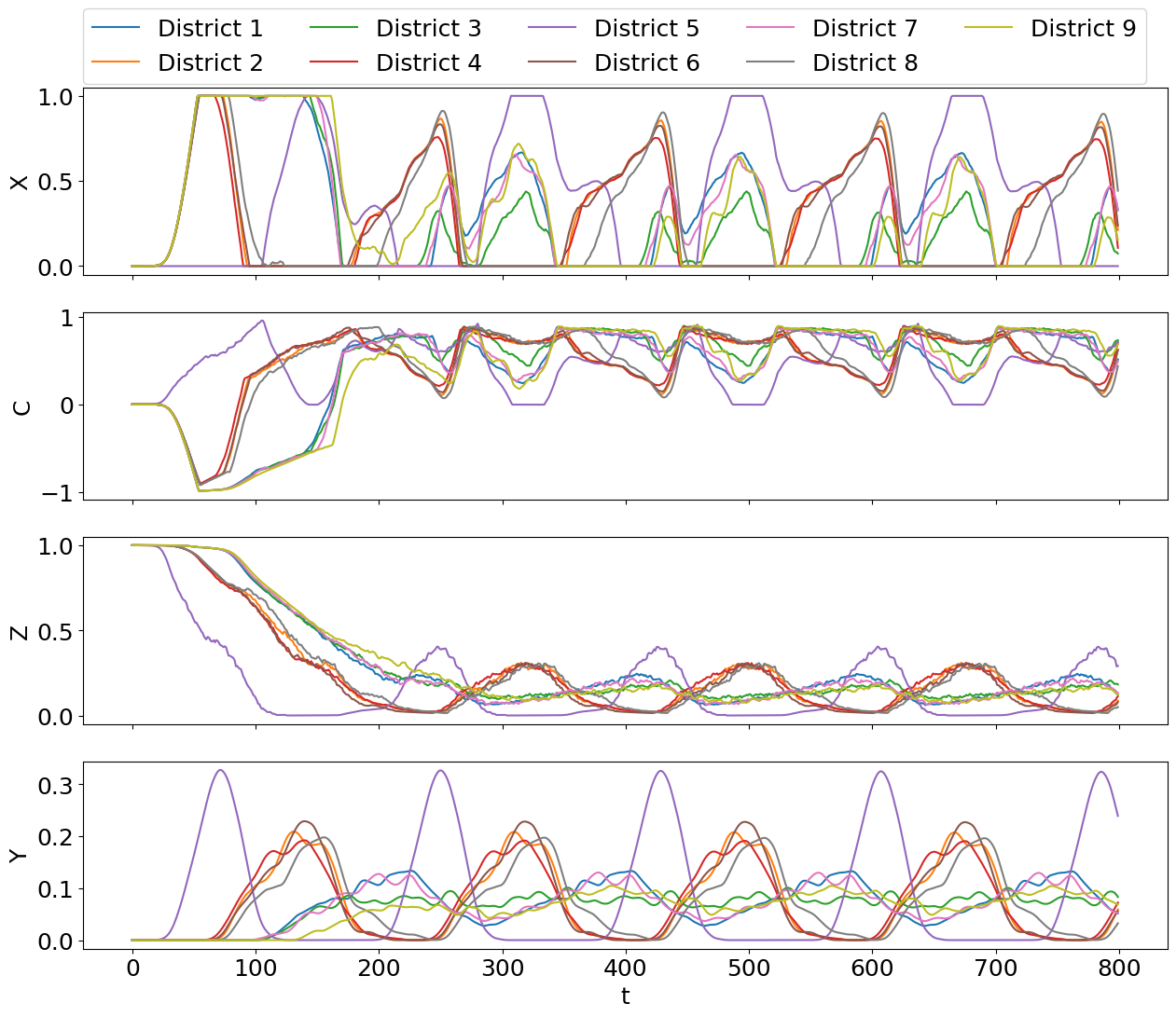}
\epsfxsize = 0.40\columnwidth \epsfysize = 0.3\textheight \epsffile{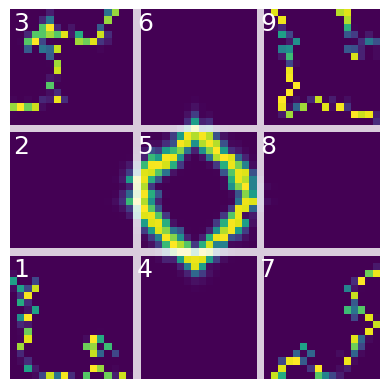}
\caption{The results of the Prohealth case, i.e, simulating health-conscious population agents. The notations are the same as in Fig. \ref{Cminus}.}
\label{WY10fig}
\end{figure}

\subsubsection*{Population agents' health concerns}

It is clear from the results of the previous subsection that the compliance characteristic of the population agents has a significant effect on their behaviour and spreading of epidemic, but what can be said of the effects of other parameters? For example if the population agents were more averse to disease spreading, would they attempt to moderate the rapid spread seen in the case of Non-compliance? To study this question we made a simulation called Prohealth, where we used otherwise the same parameters as in the case of Non-compliance, but changed the concern for health of the population agent $w^y_i = -50$, for all the agents. The results are depicted in Fig. \ref{WY10fig}, in the spreading pattern of which geographical variability in the propagation of the epidemic is enhanced. This is because there is more space between consecutive wave fronts in the cardinal directions than in the diagonal directions. The population agents are also much more willing and consistent in limiting the spread of epidemic, as can be seen in the economic activity $Z_i$ and the degree of compliance $C_i$ curves. In the former we observe relaxation to values much lower than those found in the Non-compliance simulation and the latter approaches values near the maximum value of $1$ for all the districts, with the exception of dips caused by the seasonal tightening of restrictions by the district authority agents. In the end state the economic activity $Z_i$ curves are all periodic, with the central district ($i = 5$) experiencing larger amplitudes than the edge sharing districts, which in turn have larger amplitudes than the corner sharing districts. 

In the infection rate ($Y_i$) curves of Fig. \ref{WY10fig} we see that apart from the central district ($i = 5$), where the epidemic keeps re-emerging, in the other districts the disease becomes almost endemic. Then the infection rates vary between the maximal value of about $0.2$ and a minimum value of about $0$ in the edge sharing districts and fluctuating around approximately $0.07$ in the corner sharing districts. However, in all the districts the maximal infection rates are slightly below those of the Non-compliance case. 

The restrictions imposed by the district authority agents ($X_i$) are at the beginning of epidemic rather similar to those of the Non-compliance case, but they become somewhat subdued in the end state of the simulation. Most of the districts drop their restrictions entirely, only to reintroduce them seasonally in the latter part of the time series. These seasonal adjustments seem to follow similar pattern as in the Non-compliance case, i.e. restrictions are lifted when the infection rate is low. 

\subsubsection*{District authority agents' economic concerns}

\begin{figure}[ht]
\centering
\hspace{-0.53cm}\epsfxsize = 0.45\columnwidth \epsfysize = 0.3\textheight \epsffile{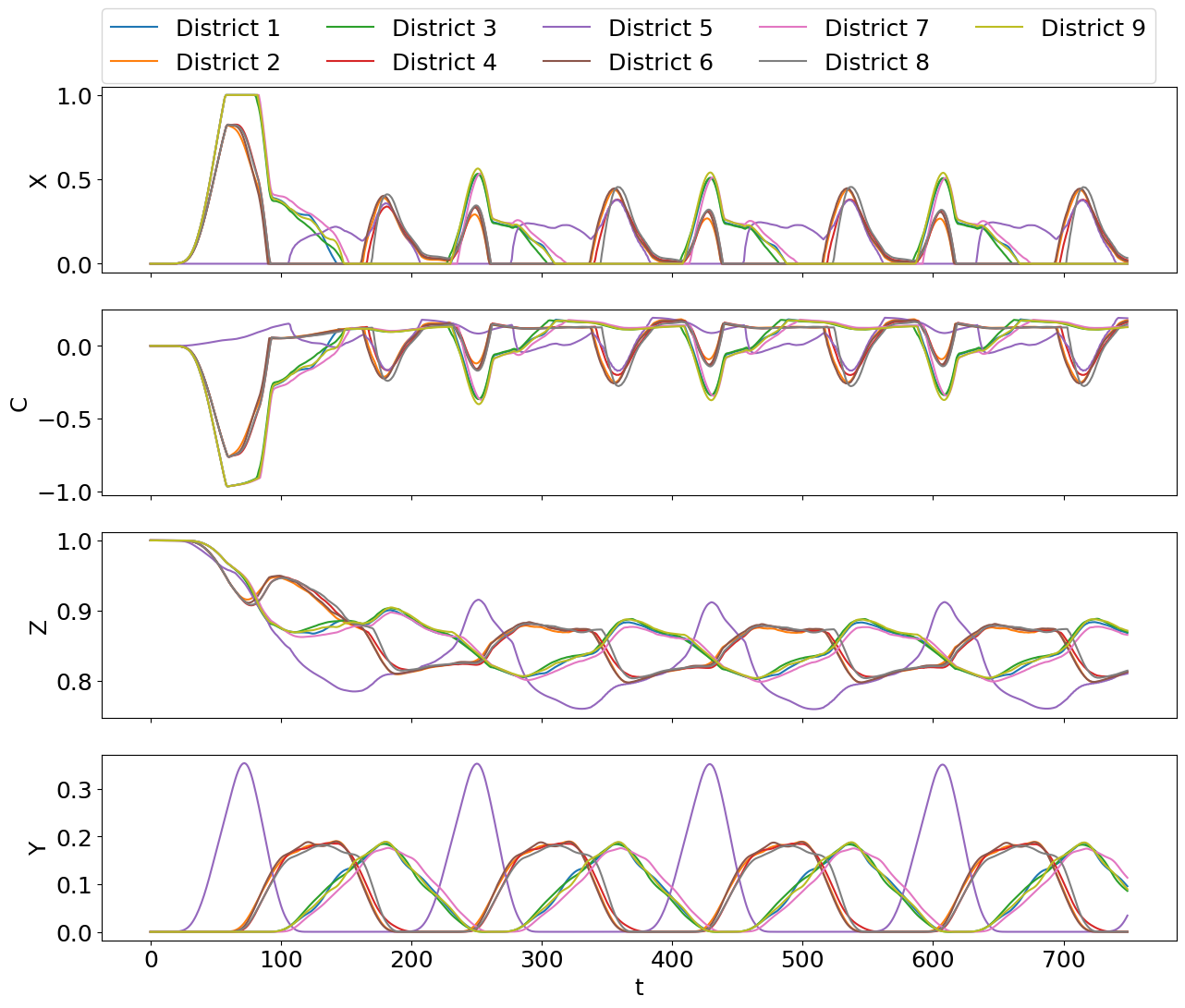}
\epsfxsize = 0.4\columnwidth \epsfysize = 0.3\textheight \epsffile{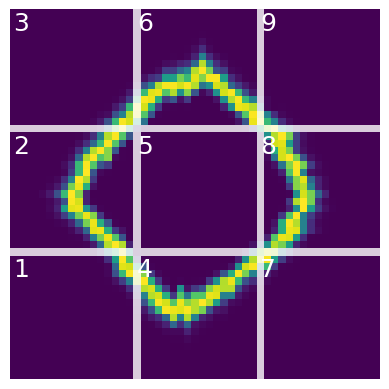}
\caption{The results of the Proeconomy case, i.e., simulating district authority agents being conscious of economy. The notations are the same as in Fig. \ref{Cminus}.}
\label{Wz10fig}
\end{figure}

Next we investigate how does the district authority agents' concerns for the economy influence the behavior of the population agents and the spread of the epidemic. We again use the Non-compliance case as our basis, but we set $W^Z_i$ of all the district authority agents to $-10$, and call it the Proeconomy simulation. The results of this simulation are depicted in Fig. \ref{Wz10fig}.

Here the spreading pattern and time-evolution of the infection rates are essentially the same as in the case of Non-compliance simulation, but there are substantial differences in the behaviour of the district authority and population agents. Most obvious of these is that the district authority agents are much less stringent when enacting restrictions. We find that while in the initial epidemic wave their restrictions ($X_i$) may reach values as high as the maximum level of $1.0$, once the later waves start rolling in, they reach up to a little over $0.5$ at most. This is naturally reflected in the degree of compliance $C_i$ curve, which indicates that in this case the population agents are often compliant with the lax restrictions by the district authority agents.

In this case a surprising observation is that the economic activity $Z_i$ is significantly affected by the change in the $W^Z_i$ parameter, as the formerly very regular curves start oscillating somewhat irregularly around some mean values. These values seem to be remarkably higher than those of the Non-Compliance case. The root cause for the weaker efforts of the population agents to contain the epidemic can be found in the third term of Eq. \ref{compUeq}. With the values of $w^x_i$ and $w^c_i$ we have chosen for Proeconomy simulation we have 
\begin{linenomath*}
\begin{equation}
\frac{w^c_i}{w^c_i + w^x_i} = \frac{2}{3},
\end{equation}
\end{linenomath*}
from which it follows 
\begin{linenomath*}
\begin{equation}
\frac{w^c_i}{w^c_i + w^x_i} |e_i| X^r_i \geq 0,
\end{equation}
\end{linenomath*}
where the equality only holds, if there are no regulations imposed by the district authority agents, i.e. $X^r_i = 0$. This, in turn, means that if the district authority agents issue lighter restrictions, the population agents will likewise lower their contributions to ward off the epidemic, which is seen in the difference between the economic activity curves of Figs. \ref{Wz10fig} and \ref{Cminus}.

\section*{Discussion}

In this study we have devised a hybrid human society and geographical area based model to investigate socio-economic aspects of epidemics. For this we took a compartmental SEIRS model with geographic aspects as the basis and implemented socio-economic dynamics by introducing local population and district authority agents that react to temporal state of epidemic spreading with Better-Than-Hypthesis or BTH-approach. In this hybrid BTH-SEIRS model the epidemic spreading takes place over the entire geographical area, divided to a grid of cells each with a population agent and collections of these cells forming a larger area district governed by its authority agent. These district authority agents impose recommendations or restrictions to lower economic activity to stem the tide of the epidemic, which the local population agents can comply with, i.e. choose to limit their economic activity to lower the transmission rate between adjacent geographical cells. One of the novelties of our model is that the agents, having information of the state of epidemic itself, can react to it to mitigate its spreading. 
For showcasing the basic properties of our model, we used a relatively rapid base spreading rate of the epidemic, and also made the mitigating efforts of epidemic by the population agents relatively effective. As a result the model exhibits rapid and regular or slow, chaotic spreading patterns as extreme cases. 

The model in its current form suggests that the willingness of the population agents to follow the restrictions implemented by the district authority agents turns out to be the most significant factor in reducing the speed at which an epidemic spreads through the whole geographical area. Indeed, the easiest way to change the geographical transmission speed of the epidemic from extremely slow to extremely fast is to give all the population agents the same value parameters with sufficiently large compliance value $w^c_i$ and switching its sign from positive to negative. The chaotic nature of the slow spreading pattern resulting from having compliant population agents, ($w^c_i > 0$), is generally reflected in the decreased infection rates and the district authority agents' restrictions, but not so much in the economic activity, which tends to go to zero almost monotonically. On the other hand the fast epidemic spreading caused by non-compliant population agents, ($w^c_i < 0$), results in periodic outbreaks of the disease and periodic responses by both the district authority and population agents.

If one interprets the parameter $w^c_i$ as measuring the adherence to social norms in general, this result is broadly in agreement with the results of \cite{GJPNPDDVCW2021}, where it was observed that culturally loose (low adherence) countries tend to suffer more of epidemic than culturally tight (high adherence) ones. Other intuitively sensible behavioural patterns arising from our model are the effects of the health and economic concerns on the spreading rates, such that higher health concerns drive up efforts to stem the epidemic, while economic concerns drive these efforts down. 

While capturing these intuitive features with the BTH-SEIRS model is encouraging, our simulations also exhibit some counter-intuitive ones. One of those is that in our simulations the district authority agents only impose restrictions when the epidemic is at low ebb and scrap them when the epidemic picks up the pace. Remembering the discussion about how this pattern emerges from the second term of Eq. \ref{dmeqgovmkII}, this feature can be understood to be a result of the competitive nature of the agents and their non-existent ability to constrain the epidemic once it has broken loose. If the epidemic is particularly serious in a district relative to others, and the effects of the local restrictions are not significant anyway, why would the authorities of the said district bother putting up any restrictions at all? This situation could change if the district authorities had a realistic chance of eliminating the disease altogether using pharmaceutical interventions as opposed to mere NPI:s. 

In the present study we chose to describe the fixed populations of geographic cells as agents that can limit the spreading of epidemic to neighbouring cells in our model. An alternative approach would have been to let the population agents be independent of their geographical locations, and letting many of them inhabit the same cells and even move around in a limited manner. This sort of setting has been used in many agent-based epidemic models \cite{WHRA2011,BDHHK2020,SBLAGS2020,KSJJ2021}, and its main advantage to our approach is more realistic modelling of human mobility: the agents can simply transmit the disease between themselves when they come close to each other. It should be noted, however, that this approach is useful in a relatively small scale, such as cities, as then agents can have realistic mobility, based on their daily schedules. However, in the present study our aim has been to look at larger-scale systems, with socio-economic dynamics at local and district level scales. 

Some other future improvements of our model could include taking into account specific NPIs, such as mandatory mask wearing, social distancing, business closures and the like, their economic effects, and improving the decision-making algorithm to handle these more realistic NPIs. One of the most significant compromises we have made 
in our model design lies in the agents' decision making algorithm. In order to arrive at the values for agents' epidemic reduction strategies $x_i$ and $X_i$ (Eqs. \ref{compUeq} and \ref{dmeqgovmkII}, respectively) we have employed a minimum effort assumption that constrains the behaviour of the agents and the space of the value parameters. It should be stressed that this choice is purely on grounds of simplicity, and that alternative, less constrained decision making algorithms are possible. One could, for example, let agents choose the utility-wise best option from randomly generated alternative strategies. Adapting the model to study the effects of vaccinations or adding economic infrastructure and businesses as economic agent to the model are other possible future prospects. 

\section*{Acknowledgements}

KK acknowledge support from EU HORIZON 2020 INFRAIA-1-2014-2015 program project (SoBigData: Social Mining and Big Data Ecosystem) No. 654024 and INFRAIA-2019-1 (SoBigData++:European Integrated Infrastructure for Social Mining and Big Data Analytics) No. 871042 as well as NordForsk Programme for Interdisciplinary Research project "The Network Dynamics of Ethnic Integration". KK also acknowledges the Visiting Fellowship at The Alan Turing Institute, UK. R.A.B. and J.E.S. acknowledge support from The National Autonomous University of Mexico (UNAM) and Alianza UCMX of the University of California (UC), through the project included in the Special Call for Binational Collaborative Projects addressing COVID-19.  M.J.K. acknowledges funding from the European Research Council (ERC) under the European Union's Horizon 2020 research and innovation programme, through project ``UniSDyn'', grant agreement n:o 818665.

\section*{Author contributions statement}

R.A.B. is the main developer of the original SEIRS model. J.E.S. modified the model to include agent-based BTH-aspects. All the authors took part in the planning of the simulation runs to be performed.
J.E.S. was responsible for practical design of these simulations, running them, and analysing their results. J.E.S. was the lead author of the manuscript, while all the co-authors contributed by writing and revising the paper. 




\section*{Data availability}

All data generated, and customised computer programs used to generate them, are available from the corresponding author on reasonable request.

\section*{Competing interests}

The authors declare no competing interests.

\bibliography{BTHgeneral}

\begin{thebibliography}{10}
\urlstyle{rm}
\expandafter\ifx\csname url\endcsname\relax
  \def\url#1{\texttt{#1}}\fi
\expandafter\ifx\csname urlprefix\endcsname\relax\def\urlprefix{URL }\fi
\expandafter\ifx\csname doiprefix\endcsname\relax\def\doiprefix{DOI: }\fi
\providecommand{\bibinfo}[2]{#2}
\providecommand{\eprint}[2][]{\url{#2}}

\bibitem{PERRA2021}
\bibinfo{author}{Perra, N.}
\newblock \bibinfo{journal}{\bibinfo{title}{Non-pharmaceutical interventions
  during the covid-19 pandemic: A review}}.
\newblock {\emph{\JournalTitle{Physics Reports}}}
  \doiprefix\url{https://doi.org/10.1016/j.physrep.2021.02.001}
  (\bibinfo{year}{2021}).

\bibitem{KK1991a}
\bibinfo{author}{Kermack, W.} \& \bibinfo{author}{McKendrick, A.}
\newblock \bibinfo{journal}{\bibinfo{title}{Contributions to the mathematical
  theory of epidemics--i. 1927}}.
\newblock {\emph{\JournalTitle{Bulletin of mathematical biology}}}
  \textbf{\bibinfo{volume}{53}}, \bibinfo{pages}{33—55},
  \doiprefix\url{10.1007/bf02464423} (\bibinfo{year}{1991}).

\bibitem{KK1991b}
\bibinfo{author}{Kermack, W.} \& \bibinfo{author}{McKendrick, A.}
\newblock \bibinfo{journal}{\bibinfo{title}{Contributions to the mathematical
  theory of epidemics—ii. the problem of endemicity}}.
\newblock {\emph{\JournalTitle{Bulletin of Mathematical Biology}}}
  \textbf{\bibinfo{volume}{53}}, \bibinfo{pages}{57--87},
  \doiprefix\url{https://doi.org/10.1016/S0092-8240(05)80041-2}
  (\bibinfo{year}{1991}).

\bibitem{KK1991c}
\bibinfo{author}{Kermack, W.} \& \bibinfo{author}{McKendrick, A.}
\newblock \bibinfo{journal}{\bibinfo{title}{Contributions to the mathematical
  theory of epidemics—iii. further studies of the problem of endemicity}}.
\newblock {\emph{\JournalTitle{Bulletin of Mathematical Biology}}}
  \textbf{\bibinfo{volume}{53}}, \bibinfo{pages}{89--118},
  \doiprefix\url{https://doi.org/10.1016/S0092-8240(05)80042-4}
  (\bibinfo{year}{1991}).

\bibitem{BVGM2013}
\bibinfo{author}{Barrio, R.}, \bibinfo{author}{Varea, C.},
  \bibinfo{author}{Govezensky, T.} \& \bibinfo{author}{José, M.}
\newblock \bibinfo{journal}{\bibinfo{title}{Modeling the geographical spread of
  influenza a(h1n1): The case of mexico}}.
\newblock {\emph{\JournalTitle{Applied Mathematical Sciences}}}
  \textbf{\bibinfo{volume}{7}}, \bibinfo{pages}{2143--2176},
  \doiprefix\url{10.12988/ams.2013.13193} (\bibinfo{year}{2013}).

\bibitem{BKHAG2020}
\bibinfo{author}{Barrio, R.~A.}, \bibinfo{author}{Kaski, K.~K.},
  \bibinfo{author}{Haraldsson, G.~G.}, \bibinfo{author}{Aspelund, T.} \&
  \bibinfo{author}{Govezensky, T.}
\newblock \bibinfo{title}{Modelling covid-19 epidemic in mexico, finland and
  iceland} (\bibinfo{year}{2020}).
\newblock \eprint{2007.10806}.

\bibitem{BGBB2020}
\bibinfo{author}{Barreiro, N.~L.}, \bibinfo{author}{Govezensky, T.},
  \bibinfo{author}{Bolcatto, P.~G.} \& \bibinfo{author}{Barrio, R.~A.}
\newblock \bibinfo{title}{Detecting infected asymptomatic cases in a stochastic
  model for spread of covid-19. the case of argentina} (\bibinfo{year}{2020}).
\newblock \eprint{2012.15209}.

\bibitem{SGGBK2017}
\bibinfo{author}{Snellman, J.~E.}, \bibinfo{author}{I{\~n}iguez, G.},
  \bibinfo{author}{Govezensky, T.}, \bibinfo{author}{Barrio, R.~A.} \&
  \bibinfo{author}{Kaski, K.~K.}
\newblock \bibinfo{journal}{\bibinfo{title}{Modelling community formation
  driven by the status of individual in a society}}.
\newblock {\emph{\JournalTitle{Journal of Complex Networks}}}
  \textbf{\bibinfo{volume}{5}}, \bibinfo{pages}{817--838},
  \doiprefix\url{10.1093/comnet/cnx009} (\bibinfo{year}{2017}).
\newblock
  \eprint{/oup/backfile/content_public/journal/comnet/5/6/10.1093_comnet_cnx009/5/cnx009.pdf}.

\bibitem{SGKBK2018}
\bibinfo{author}{Snellman, J.~E.}, \bibinfo{author}{I{\~n}iguez, G.},
  \bibinfo{author}{Kert{\'e}sz, J.}, \bibinfo{author}{Barrio, R.~A.} \&
  \bibinfo{author}{Kaski, K.~K.}
\newblock \bibinfo{journal}{\bibinfo{title}{Status maximization as a source of
  fairness in a networked dictator game}}.
\newblock {\emph{\JournalTitle{Journal of Complex Networks}}}
  \bibinfo{pages}{cny022}, \doiprefix\url{10.1093/comnet/cny022}
  (\bibinfo{year}{2018}).
\newblock
  \eprint{/oup/backfile/content_public/journal/comnet/pap/10.1093_comnet_cny022/1/cny022.pdf}.

\bibitem{SBK2021}
\bibinfo{author}{Snellman, J.~E.}, \bibinfo{author}{Barrio, R.~A.} \&
  \bibinfo{author}{Kaski, K.~K.}
\newblock \bibinfo{journal}{\bibinfo{title}{Social structure formation in a
  network of agents playing a hybrid of ultimatum and dictator games}}.
\newblock {\emph{\JournalTitle{Physica A: Statistical Mechanics and its
  Applications}}} \textbf{\bibinfo{volume}{561}}, \bibinfo{pages}{125257},
  \doiprefix\url{https://doi.org/10.1016/j.physa.2020.125257}
  (\bibinfo{year}{2021}).

\bibitem{adler}
\bibinfo{author}{Adler, A.}
\newblock \emph{\bibinfo{title}{The Practice and Theory of Individual
  Psychology}} (\bibinfo{publisher}{Routledge, Trench and Trubner \& Co, Ltd},
  \bibinfo{year}{1924}).
\newblock \bibinfo{note}{Reprint. Abingdon: Routledge (1999)}.

\bibitem{AHH2015}
\bibinfo{author}{Anderson, C.}, \bibinfo{author}{Hildreth, J. A.~D.} \&
  \bibinfo{author}{Howland, L.}
\newblock \bibinfo{journal}{\bibinfo{title}{Is the desire for status a
  fundamental human motive? a review of the empirical literature}}.
\newblock {\emph{\JournalTitle{Psychological Bulletin}}}
  \textbf{\bibinfo{volume}{141}}, \bibinfo{pages}{574--601}
  (\bibinfo{year}{2015}).

\bibitem{ZTCB2008}
\bibinfo{author}{Zink, C.~F.} \emph{et~al.}
\newblock \bibinfo{journal}{\bibinfo{title}{Know your place: Neural processing
  of social hierarchy in humans}}.
\newblock {\emph{\JournalTitle{Neuron}}} \textbf{\bibinfo{volume}{58}},
  \bibinfo{pages}{273--283} (\bibinfo{year}{2008}).

\bibitem{ISS2008}
\bibinfo{author}{Izuma, K.}, \bibinfo{author}{Saito, D.~N.} \&
  \bibinfo{author}{Sadato, N.}
\newblock \bibinfo{journal}{\bibinfo{title}{Processing of social and monetary
  rewards in the human striatum}}.
\newblock {\emph{\JournalTitle{Neuron}}} \textbf{\bibinfo{volume}{58}},
  \bibinfo{pages}{284--294} (\bibinfo{year}{2008}).

\bibitem{KMD2012}
\bibinfo{author}{Kumaran, D.}, \bibinfo{author}{Melo, H.~L.} \&
  \bibinfo{author}{Duzel, E.}
\newblock \bibinfo{journal}{\bibinfo{title}{The emergence and representation of
  knowledge about social and nonsocial hierarchies}}.
\newblock {\emph{\JournalTitle{Neuron}}} \textbf{\bibinfo{volume}{76}},
  \bibinfo{pages}{653 -- 666},
  \doiprefix\url{https://doi.org/10.1016/j.neuron.2012.09.035}
  (\bibinfo{year}{2012}).

\bibitem{UP2014}
\bibinfo{author}{Utevsky, A.~V.} \& \bibinfo{author}{Platt, M.~L.}
\newblock \bibinfo{journal}{\bibinfo{title}{Status and the brain}}.
\newblock {\emph{\JournalTitle{PLOS Biology}}} \textbf{\bibinfo{volume}{12}},
  \bibinfo{pages}{1--4}, \doiprefix\url{10.1371/journal.pbio.1001941}
  (\bibinfo{year}{2014}).

\bibitem{IT2020}
\bibinfo{author}{Inoue, H.} \& \bibinfo{author}{Todo, Y.}
\newblock \bibinfo{journal}{\bibinfo{title}{The propagation of economic impacts
  through supply chains: The case of a mega-city lockdown to prevent the spread
  of covid-19}}.
\newblock {\emph{\JournalTitle{PLOS ONE}}} \textbf{\bibinfo{volume}{15}},
  \bibinfo{pages}{1--10}, \doiprefix\url{10.1371/journal.pone.0239251}
  (\bibinfo{year}{2020}).

\bibitem{IMT2021}
\bibinfo{author}{Hiroyasu, I.}, \bibinfo{author}{Yohsuke, M.} \&
  \bibinfo{author}{Yasuyuki, T.}
\newblock \bibinfo{title}{{Do Economic Effects of the Anti-COVID-19 Lockdowns
  in Different Regions Interact through Supply Chains?}}
\newblock \bibinfo{type}{Discussion papers} \bibinfo{number}{21001},
  \bibinfo{institution}{Research Institute of Economy, Trade and Industry
  (RIETI)} (\bibinfo{year}{2021}).

\bibitem{CLV2020}
\bibinfo{author}{Correia, S.}, \bibinfo{author}{Luck, S.} \&
  \bibinfo{author}{Verner, E.}
\newblock \bibinfo{journal}{\bibinfo{title}{Pandemics depress the economy,
  public health interventions do not: Evidence from the 1918 flu}}.
\newblock {\emph{\JournalTitle{Available at SSRN:
  https://ssrn.com/abstract=3561560 or
  http://dx.doi.org/10.2139/ssrn.3561560}}}  (\bibinfo{year}{(June 5, 2020).}).

\bibitem{BDHHK2020}
\bibinfo{author}{Basurto, A.}, \bibinfo{author}{Dawid, H.},
  \bibinfo{author}{Harting, P.}, \bibinfo{author}{Hepp, J.} \&
  \bibinfo{author}{Kohlweyer, D.}
\newblock \bibinfo{journal}{\bibinfo{title}{Economic and epidemic implications
  of virus containment policies: Insights from agent-based simulations}}.
\newblock {\emph{\JournalTitle{Bielefeld Working Papers in Economics and
  Management}}}  (\bibinfo{year}{(June 24, 2020).}).

\bibitem{SBLAGS2020}
\bibinfo{author}{Silva, P.~C.} \emph{et~al.}
\newblock \bibinfo{journal}{\bibinfo{title}{Covid-abs: An agent-based model of
  covid-19 epidemic to simulate health and economic effects of social
  distancing interventions}}.
\newblock {\emph{\JournalTitle{Chaos, Solitons \& Fractals}}}
  \textbf{\bibinfo{volume}{139}}, \bibinfo{pages}{110088},
  \doiprefix\url{https://doi.org/10.1016/j.chaos.2020.110088}
  (\bibinfo{year}{2020}).

\bibitem{KSJJ2021}
\bibinfo{author}{Koehler, M.}, \bibinfo{author}{Slater, D.~M.},
  \bibinfo{author}{Jacyna, G.} \& \bibinfo{author}{Thompson, J.~R.}
\newblock \bibinfo{journal}{\bibinfo{title}{Modeling covid-19 for lifting
  non-pharmaceutical interventions}}.
\newblock {\emph{\JournalTitle{Journal of Artificial Societies and Social
  Simulation}}} \textbf{\bibinfo{volume}{24}}, \bibinfo{pages}{9},
  \doiprefix\url{10.18564/jasss.4585} (\bibinfo{year}{2021}).

\bibitem{GJPNPDDVCW2021}
\bibinfo{author}{Gelfand, M.~J.} \emph{et~al.}
\newblock \bibinfo{journal}{\bibinfo{title}{The relationship between cultural
  tightness-looseness and covid-19 cases and deaths: a global analysis}}.
\newblock {\emph{\JournalTitle{The Lancet Planetary Health}}}
  \doiprefix\url{10.1016/S2542-5196(20)30301-6} (\bibinfo{year}{2021/02/01}).

\bibitem{WHRA2011}
\bibinfo{author}{Williams, A.~D.}, \bibinfo{author}{Hall, I.~M.},
  \bibinfo{author}{Rubin, G.~J.}, \bibinfo{author}{Amlôt, R.} \&
  \bibinfo{author}{Leach, S.}
\newblock \bibinfo{journal}{\bibinfo{title}{An individual-based simulation of
  pneumonic plague transmission following an outbreak and the significance of
  intervention compliance}}.
\newblock {\emph{\JournalTitle{Epidemics}}} \textbf{\bibinfo{volume}{3(2)}},
  \bibinfo{pages}{95--102}, \doiprefix\url{doi:10.1016/j.epidem.2011.03.001}
  (\bibinfo{year}{2011}).

\end{thebibliography}




\end{document}